\documentstyle[preprint,aps]{revtex}
\tighten
\begin{document}
\draft
\title{
Precursors of Mott insulator in modulated quantum wires
       }
\author{ Arkadi A. Odintsov$^{1,2}$, Yasuhiro Tokura$^{2}$, 
	 Seigo Tarucha$^{2}$}
\address{
$^{1}$Department of Applied Physics,
Delft University of Technology,
2628 CJ Delft, The Netherlands \\
and Nuclear Physics Institute, Moscow
State University, Moscow 119899 GSP, Russia
	}
\address{$^{2}$NTT Basic Research Laboratories
3-1 Morinosato-Wakamiya, Atsugi-shi 243-01, Japan}

\maketitle
\begin{abstract}
We investigate the transport of interacting electrons through  
single-mode quantum wires whose parameters are periodically modulated on 
the scale of the electronic Fermi wave length.
The Umklapp and backscattering of electrons
can be described in terms of non-uniform quantum 
sine-Gordon-like models which also incorporate the effects 
of electronic reservoirs (electrodes) adiabatically coupled to the wire.
We concentrate on weak Umklapp scattering 
and analyze the precursors of the Mott transition.
At half-filling the temperature dependence of 
the extra resistance $\Delta R = R - \pi \hbar/e^2$ 
of a modulated quantum wire of length $L$
changes from the interaction-dependent "bulk" power-law 
$\Delta R \propto T^{4K_\rho-3}$ 
at high temperatures, $T \gg v_\rho/L$,
to the universal $\Delta R \propto T^2$ behavior 
at low temperatures, $T \ll v_\rho/L$.
Away from half-filling the "bulk" results 
are qualitatively incorrect even at high temperatures
$v_\rho/L \ll T \ll T^{*}$ 
despite the electron coherence in the wire is absent
in this regime.
\end{abstract} 


Since its discovery in 1949 the Mott transition (MT) \cite{Mott}
remains in focus of intensive investigations. 
Much progress has been achieved in the theory of
the MT in one spatial dimension (see reviews \cite{Voit,Schulz,Emery}).
Recent breakthroughs in technology  
have opened at least two intriguing opportunities 
for experimental observation of the MT in artificially fabricated and 
naturally grown one-dimensional (1D) conductors.

First, it became possible to
fabricate long and clean quantum wires \cite{Tarucha}.
This system has been successfully used for the detection of
non-Fermi liquid behavior of interacting electrons in 1D \cite{Tarucha,Yacobi}.
It should be technologically possible
to modulate the electrostatic potential along the wire 
with the period $a=2\pi/G$ of the order of the electronic 
Fermi wave length using selective wet etching of the donor layer.
By varying the concentration of electrons with an additional gate 
one can study the effects of the electron backscattering and Umklapp 
scattering on transport through modulated quantum wire (MQW). In an infinite system 
the electron backscattering leads to the opening of the gap at the boundary 
of the Brillouin zone,  $G = 2 k_F$.  More interestingly, at half-filling,  
$G = 4k_{F}$, the Mott gap is formed \cite{comm1} 
due to the Umklapp scattering which 
occurs only in the interacting systems \cite{Voit,Schulz,Emery}. 
By changing the position of the Fermi level with respect to the gaps 
one can effectively control electronic transport. 

Second, drastic progress has been recently achieved 
in the synthesis of single wall carbon nanotubes \cite{Thess}.
Coherent electron transport and single-electron effects in this system
have been demonstrated in recent experiments \cite{Tans}.
Theoretically, low-energy properties of "armchair" nanotubes  
can be described by a two-chain Hubbard model at 
half-filling \cite{Fisher}.
The Umklapp scattering causes these (otherwise metallic)
nanotubes to experience a MT at low temperatures \cite{Krotov,Fisher}. 

Previous research has addressed the effect of Umklapp 
scattering on the transport in uniform (and formally infinite)
1D systems \cite{Giamarchi,Krotov,Fisher}.

In a realistic experimental layout the system 
(MQW or carbon nanotube)
is connected to external electric contacts 
which can be considered as reservoirs of non-interacting electrons.
The presence of the reservoirs makes the system {\it non-uniform}.
This has important consequences for the transport 
through clean \cite{Maslov} and dirty \cite{dirty}  quantum wires.
The investigation of the effects of Umklapp scattering and backscattering 
of electrons on transport through realistic MQWs
forms the focus of this work.
 
\noindent
{\it Umklapp scattering},  $G \simeq 4 k_F$. 
We consider a single-channel MQW
adiabatically coupled to two perfect non-interacting 1D leads 
which model electronic reservoirs \cite{Maslov} (see insert of Fig.~1).
At low energies $E \ll E_{F}$ the system can be treated within 
the bosonization formalism \cite{Voit,Schulz}. To describe the transport 
it is enough to consider the charge part 
$H = H_\rho+H_{U}$ of the bosonized Hamiltonian, 
which is decoupled from the spin part. 
The charge part contains the standard Tomonaga-Luttinger term
($\hbar = 1$),
\begin{equation}
H_\rho = \int_{-\infty}^{\infty}
\frac{dx}{\pi}
\left\{
	\frac{v_{\rho}(x)}{K_{\rho}(x)} (\nabla \Theta_{\rho})^2 + 
	v_{\rho}(x) K_{\rho}(x) (\nabla \phi_{\rho})^2
\right\}
\label{H_rho}
\end{equation}
associated with the forward scattering of electrons
and the nonlinear term
\begin{equation}
H_U = \int_{-\infty}^{\infty}
dx U(x) \cos 4 \Theta_{\rho}
\label{H_u}
\end{equation}
describing the Umklapp scattering.
Here  $\theta_\rho$ and $\phi_\rho$ are bosonic fields
satisfying the commutation relation
$[\theta _\rho (x), \phi _\rho (x')] =  (i \pi /4) \mbox{sign} (x-x')$,
$v_\rho$ is the velocity of charge excitations,
and $K_\rho$ is a standard interaction parameter of the Tomonaga-Luttinger 
model ($K_\rho=1$ for non-interacting electrons). 
The amplitude $U$ of the Umklapp scattering is proportional to 
the $2k_{F}$-Fourier component $V(2k_{F})$ of the electron-electron interaction.
In particular, for a weak periodic potential $W(x) = W_{0} \cos Gx$,
$W_{0} \ll E_{F}$ we obtain $U = V(2k_{F}) W_{0} / 2 \pi v_{F}a$.

We will assume that the Umklapp scattering as well as 
the Coulomb interaction of electrons occur only in 
the MQW ($|x| < L/2$)
which is characterized by position independent parameters 
$(K_\rho(x),v_\rho(x),U(x)) = (K,v_w,U)$ in (\ref{H_rho}), (\ref{H_u}).
The parameters change stepwise at $x = \pm L/2$
acquiring non-interacting values $(1,v_l,0)$
in the leads ($|x| > L/2$) \cite{Maslov}.
Let us note, that despite the assumption that the wire 
is coupled to the leads via adiabatic contacts with size 
$R \gg 2\pi/k_{F}$
the use of stepwise approximation is legitimate at 
low energies, $E \ll v_F/R \ll E_F$.

To evaluate the current 
$I = 2e \langle \dot{\theta_\rho} \rangle / \pi$
through the wire we consider the Heisenberg equations 
for $\dot{\theta_\rho}$ and $\dot{\phi_\rho}$. 
The equations should be
supplemented with the boundary conditions \cite{egger}
\begin{equation}
 \nabla \langle \theta_\rho \pm \phi_\rho \rangle 
	|_{x \to \mp \infty} = \mu_{\pm}/v_{l}¥
\label{bc}
\end{equation}
which reflect the fact that the chemical potential 
$\mu_+$ ( $\mu_-$) 
of the left (right) reservoir determines an excess density 
$\rho_+$ ( $\rho_-$) 
of rightgoing (leftgoing) electrons in the left (right) lead, 
$\rho_{\pm} = \mu_\pm/\pi v$
($\rho_{\pm} = 0$ corresponds to half filling, $G = 4k_F$).

In what follows we will concentrate on precursors 
of the MT  due to weak Umklapp scattering which can be treated perturbatively.
We decompose the fields $\theta_\rho$ and $\phi_\rho$ 
into classical parts $\theta_{cl}$, $\phi_{cl}$ (c-numbers)
and fluctuations $\hat{\theta}$, $\hat{\phi}$. 
In the absence of Umklapp scattering 
($U = 0$) the solution of the Heisenberg equations
satisfying the boundary conditions (\ref{bc}) has the form
\begin{equation}
\theta_{cl}^{(0)} (x,t) = \frac{q}{4} x - \frac{eV}{2} t
\; \mbox{ for } |x| < \frac{L}{2},
\label{theta_cl}
\end{equation}
where $eV = \mu_+ - \mu_-$ is the DC voltage applied, 
and $q = 2K(\mu_+ + \mu_-)/v_w =  4k_F - G$
characterizes a deviation of the electron density from half filling.
The current following from this solution corresponds to the Landauer formula, 
$I^{(0)} = (2e^2/h) V$.

The correction $\Delta I$ to the current $I^{(0)}$ 
due to Umklapp processes arises to the second order in 
the scattering amplitude $U$. It can be found by expressing the difference of
electronic densities 
$
\langle  \nabla \theta_\rho(-\infty) 
	- \nabla \theta_\rho(\infty) 
\rangle
$
at the ends of the wire from the Heisenberg equation for 
$\dot{\phi}_\rho$ and substituting the result into the
boundary condition (\ref{bc}).
After some algebra we obtain \cite{Averin},
\begin{equation}
\Delta I =
-2eU^2 \int_{-L/2}^{L/2} \frac{dx}{\pi}  \frac{dx'}{\pi} 
\int_{0}^{\infty} dt
\sin \left[ q(x-x') - \Omega t \right] 
\rm{Im} G(x,x';t),
\label{deltaI}
\end{equation}
with $\Omega = 2eV$ and
$
G(x,x';t) = 
\langle 
	e^{4i\hat{\theta} (x,t)}
	e^{-4i\hat{\theta} (x',0)}
\rangle,
$
where the average is taken over equilibrium fluctuations of the
field $\hat{\theta} (x,t)$ described by Eq. (\ref{H_rho}).
The function $G(x,x';t)$ is given by,
\begin{equation}
G(x,x';t) =
\exp 
\left\{ -8
\langle
	\hat{\theta} \hat{\theta} 
	- 2 \hat{\theta} \hat{\theta}' 
	+  \hat{\theta}' \hat{\theta}'
\rangle 
\right\},
\label{G}
\end{equation}
where $\hat{\theta} = \hat{\theta}(x,t)$ and
	$\hat{\theta}' = \hat{\theta}(x',0)$.
The correlators (\ref{G}) are related to the imaginary
part of the retarded Green's function $D_\omega^{(R)}(x,x')$
via the fluctuation-dissipation theorem ($k_{B}=1$),
\begin{equation}
\langle
	\hat{\theta} \hat{\theta}' 
\rangle
 =
- \int \frac{d\omega}{2\pi} e^{-i\omega t} 
	\left( \coth \frac{\omega}{2T} + 1 \right)
 	\rm{Im} D_\omega^{(R)}(x,x').
\label{fdt}
\end{equation}

To evaluate the retarded Green's function
$
D_\omega^{(R)}(x,x')
$
we write down the Euclidian Lagrangian corresponding to the
Hamiltonian (\ref{H_rho}), solve the Euler-Lagrange
equation for the Matsubara Green's function 
$D_{\omega}^{(M)}(x,x')$, \cite{Maslov}
and continue the latter to real frequencies,
$D_\omega^{(R)}(x,x') = D_{-i\omega+\epsilon}^{(M)}(x,x')$,
$\epsilon \to +0$. 
This gives the following expression for the spectral function,
\begin{equation}
\rm{Im}  D_\omega^{(R)}(x,x')  =
- \frac{\pi K}{4\omega}
\frac{(K_+^4-K_-^4) \cos p(x-x') + 8K K_+ K_- \cos p(x+x') \cos pL}
     {K_+^4 - 2 K_+^2 K_-^2 \cos 2pL + K_-^4},
\label{ImD}
\end{equation}
where $K_\pm = 1 \pm K$ and $p =\omega/v_w$.

Although the results (\ref{deltaI})-(\ref{ImD})  
are generally valid for non-linear transport,
in this paper we will concentrate on the linear response regime.
The latter can be conveniently described in terms of the extra resistance
$\Delta R = -(h/2e^2)^2 \lim_{V \to 0} \Delta I/V$
of the quantum wire due to the Umklapp scattering.

In order to elucidate the physics of the problem 
we start from the "non-interacting" case, $K = 1$,
where  analytical results can readily be obtained
(we imply that there is still a short range
component  of the Coulomb interaction,
$U \propto V(2k_F) \ne 0$). 
The extra resistance of MQW is given by
\begin{equation}
\Delta R  =
R_0 \int_0^l d\xi (l-\xi) \cos (\tilde{q}\xi) 
\frac{\xi \coth\xi -1}{\sinh^2 \xi},
\label{deltaR}
\end{equation}
where  
$R_0 = (2 \pi U^2 v_w^2/ \omega_c^4) h/2e^2$ 
is the scale of resistance, 
$l=2\pi T L/v_w$ 
is the dimensionless temperature,
$\tilde{q} = qv_{w}/2\pi T$
parameterizes the deviation from half filling,
and $\omega_c \sim E_F$ is the high-frequency cutoff 
in the integration (\ref{fdt}).

We start from the case of half filling ($q=0$).   
At low temperatures $T \ll v_w/L$ we obtain  
$\Delta R = (2 \pi^2 R_0/3) L^2T^2/v_w^2$.
The quadratic dependence on the length $L$ signals that 
the amplitudes of the Umklapp scattering sum up coherently 
along the MQW. At high temperatures $T \gg v_w/L$ the coherence
is lost and the extra resistance is proportional to the length of the wire,
$\Delta R = \pi R_0 LT/v_w$. 
Note that the temperature dependence of the extra resistance changes 
from quadratic to linear with increasing temperature.

Away from half filling ($q \ne 0$) the system displays three 
different regimes depending on the temperature (see Fig.~2).
At low temperatures  $T \ll v_w/L$ the interference effects show up
in the oscillatory dependence of the resistance on the mismatch parameter $q$,
\begin{equation}
\Delta R  =
\frac{4 \pi^2 R_0 T^2}{3 v_w^2 q^2}
(1-\cos qL).
\label{deltaR_low}
\end{equation}
In the intermediate temperature range $v_w/L \ll T \ll T^{*}$ 
(the parameter $T^{*}$ will be determined below) the 
oscillations disappear and the extra resistance is given by 
Eq.~(\ref{deltaR_low}) with $\cos qL = 0$. 
It might be surprising that the extra resistance does not depend on the length
of the wire in the regime when the electron coherence in the wire is destroyed by 
thermal fluctuations.

At even higher temperatures $T \gg T^{*}$ 
the extra resistance shows thermally activated behavior,
\begin{equation}
\Delta R  =
 \frac{\pi R_0 L \Delta_q^2}{2 v_w T} \frac{1}{\cosh(\Delta_q/T) - 1},
\label{deltaR_high}
\end{equation}
with $\Delta_q = v_w q/2$, in agreement with the result
for the conductance of an infinite system \cite{Giamarchi}. 
By comparing Eq.~(\ref{deltaR_high}) with the result for 
intermediate temperatures we obtain 
$T^{*} = 2\Delta_q/\ln(\Delta_q/(v_w/L))$.
Therefore, the "bulk" result (\ref{deltaR_high}) becomes
valid only at surprisingly high temperatures.

Now we turn to the interacting case ($K \ne 1$).
The temperature dependence of the resistance
is determined by the behavior of the correlator 
$G(x,x';t)$, Eq.~(\ref{G}) 
(or $D^{(R)}(x,x';t)$, Eq.~(\ref{ImD}))
at the time scale $\sim 1/T$. 
At low temperatures $T \ll v_w/L$ this time scale
corresponds to low frequencies $\omega \ll v_w/L$
at which the spectral function (\ref{ImD}) is determined by
non-interacting electrons in the leads,
$\rm{Im} D_\omega^{(R)}(x,x')  = - \pi/4\omega$.
For this reason, the extra resistance is proportional to 
$T^2$ as in the non-interacting case, see Figs.~1,2.

At high temperatures  $T \gg v_w/L$  the behavior of the
correlator $G(x,x';t)$ (\ref{G}) on the time scale $1/T$
can be evaluated by averaging the spectral function 
(\ref{ImD}) over fast oscillations with frequencies $\sim v_w/L$.
The averaged spectral function has a simple form,
\begin{equation}
\rm{Im} \bar{D}_\omega^{(R)}(x,x')  =
- \frac{\pi K}{4\omega} \cos [\omega(x-x')/v_w].
\label{ImD_appr}
\end{equation}

By substituting the approximation (\ref{ImD_appr}) into 
(\ref{deltaI})-(\ref{fdt}) we obtain the power law behavior of 
the extra resistance at high temperatures $T \gg v_w/L$ and 
half-filling ($q=0$),
\begin{equation}
\Delta R =  \frac{\alpha U^2 v_w L}{e^2 \omega_c^{4K}} T^{4K-3},
\label{dR_int}
\end{equation} 
where $\alpha$ is a non-universal numerical factor (Fig.~1).

The high temperature result (\ref{dR_int}) agrees with
the lowest order perturbative calculation of 
the DC conductivity of an infinite system \cite{Giamarchi}.
On the other hand, it is well-known that in an infinite system 
at half filling the Mott gap $\Delta_{M}$ is formed for an arbitrarily small 
amplitude $U$ of Umklapp scattering provided that the Coulomb
interaction is repulsive \cite{Voit}. At low temperatures 
$T \sim \Delta_{M}$ the perturbative result \cite{Giamarchi} 
breaks down and the resistance starts to increase exponentially. 
Our results are valid at arbitrarily low temperatures for sufficiently short 
wires, $v_{w}/L \gg \Delta_{M}$. 

Note that unlike the case of 
the Luttinger liquid with impurities \cite{dirty,FisGl} 
the high temperature behavior of the extra resistance (\ref{dR_int}) 
gives no direct indication of the true low temperature properties of the system 
(the extra resistance (\ref{dR_int}) decreases 
with decreasing temperature for $K_{\rho} > 3/4$,
despite the formation of the gap). 
On the other hand, the fact that the extra resistance (\ref{dR_int}) 
for a repulsively interacting system ($K_{\rho} < 1$) decreases somewhat slower 
than in the non-interacting case or even increases with 
decreasing temperature can be interpreted as 
a precursor of the Mott transition. 

The temperature dependence of the extra resistance 
away from half-filling ($q \ne 0$)  is presented in Fig.~2.
At high temperatures $T \gg \max(\Delta_{q}, v_{w}/L)$
this dependence obeys the power law (\ref{dR_int}).
Note that also at intermediate temperatures $v_{w}/L \ll T \ll \Delta_{q}$
the $\Delta R (T)$  dependence  is clearly affected by the interaction.
In particular, for strong enough interaction 
the resistance shows an {\it anomalous enhancement} with decreasing temperature
(see curve for $K = 0.1$ in Fig.~2).
This enhancement is a reminiscence of the corresponding effect at 
high temperatures $T \gg  \Delta_{q}$.
The variations of the extra resistance as a function of the mismatch parameter $q$ 
disappear at higher temperatures for stronger Coulomb interaction.
This can be interpreted as an enhancement of quantum interference effects
in the interacting system.

\noindent
{\it Electron backscattering}, $G \simeq 2k_F$.
Apart from the Umklapp scattering (\ref{H_u}),
the backscattering of electrons is described by the Hamiltonian
\begin{equation}
H_b = \int_{-\infty}^{\infty}
dx U(x) \cos 2 \Theta_{\rho} \cos 2 \Theta_{\sigma}
\label{H_b}
\end{equation}
which couples the charge ($\rho$) and spin ($\sigma$) degrees of 
freedom \cite{details}.
The backscattering current is given by the formula analogous to
(\ref{deltaI}) with 
\begin{equation}
G(x,x';t) =
\exp 
\left\{ -2 \Sigma_{j=\rho, \sigma}
\langle
	\hat{\theta}_j \hat{\theta}_j 
	- 2 \hat{\theta}_j \hat{\theta}'_j 
	+  \hat{\theta}'_j \hat{\theta}'_j
\rangle 
\right\},
\label{Gbragg}
\end{equation}
$q= 2k_F - G$, and $\Omega = eV$.
At filling one ($G = 2k_{F}$) and low temperatures $T \ll v_{w}/L$
the extra resistance shows no dependence on temperature and 
on the interaction strengths. At high temperatures $T \gg v_{w}/L$
we obtain $\Delta R \propto T^{K_{\rho}+K_{\sigma}-3}$,
$K_{\sigma}$ being the interaction parameter in the spin channel.

In conclusion, we have analyzed the precursors of the Mott transition 
observable in transport through modulated quantum wires
connected to electronic reservoirs. 
Known results for an infinite ("bulk") system are valid 
only at rather high temperatures $T \gg \max(v_{w}/L, T^{*})$.
In particular, at half filling the "bulk" power law dependence  of the 
extra resistance, $ \Delta R \propto T^{4K-3}$ at high  temperatures 
$T \gg v_{w}/L$ crosses over to the universal low temperature ($T \ll v_{w}/L$)
behavior, $\Delta R \propto T^{2}$,  which does not depend on the 
interaction in the wire. 
Surprisingly, away from half-filling the "bulk" results 
fail to be valid at the intermediate temperature range 
$v_{w}/L \ll T \ll T^{*}$ 
despite the electron coherence in the wire is lost in this regime. 
An anomalous enhancement of the Umklapp scattering with decreasing 
temperature for strongly interacting system in this regime should be 
contrasted to its exponential suppression expected from the "bulk" theory.
Our results are expected to be relevant also for armchair carbon 
nanotubes albeit the presence of the two energy bands in this system
should be carefully taken into account.

We would like to thank D.V. Averin, G.E.W. Bauer and Yu.V. Nazarov 
for a set of useful discussions. 
The financial support of the European Community
through HCM ERB-CHBI-CT94-1474
and Dutch Foundation for Fundamental Research on Matter (FOM)
is gratefully acknowledged. 
This work is also a part of INTAS-RFBR 95-1305.
One of us (A.O.) acknowledges
the kind hospitality at the NTT Basic Research Laboratories.

\begin{figure}
\caption{The extra resistance of modulated quantum wire at half 
filling ($q=0$) as a function of temperature. $K = 1, 0.75, 0.5, 0.25$
for the curves from top to bottom at high temperatures. 
Insert: layout of the system}
\label{fig1}
\end{figure}
\begin{figure}

\caption{The same as Fig.~1, but away from half filling. Different 
curves in each family correspond to $qL/\pi=30 \ldots 31$.
$K = 1, 0.5, 0.25, 0.1$
for the curves from top to bottom at high temperatures.
Dash-dotted lines correspond to asymtotics at intermediate and high 
temperatures (see Eqs.~10, 11 and the text in between).}
\label{fig2}
\end{figure}


\begin{references}
\bibitem{comm1} Precise conditions for this are given e.g.
in Ref. \cite{Emery}.
\bibitem{details} More detailed consideration of this case will be presented elsewhere.
\end{references}
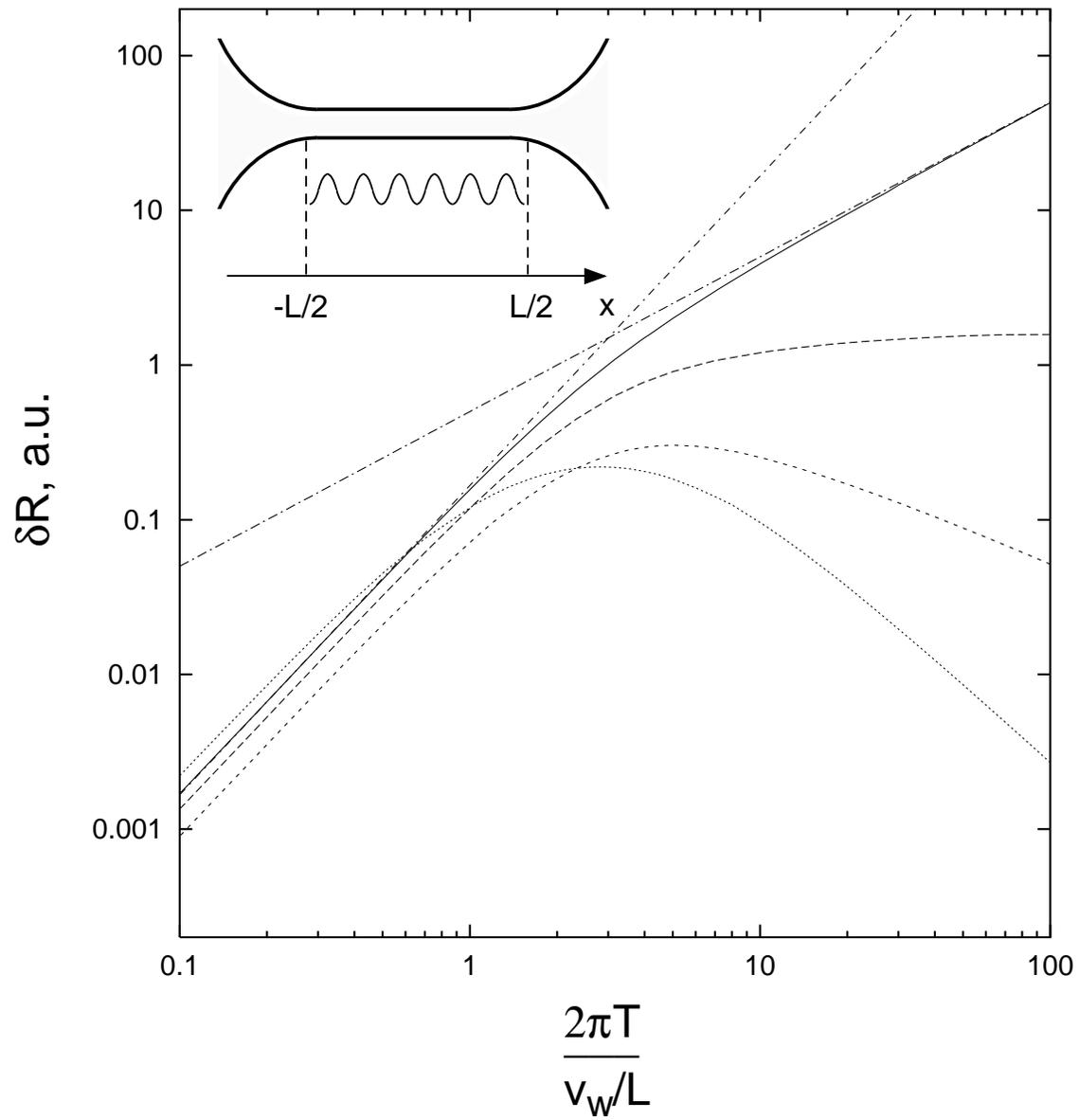
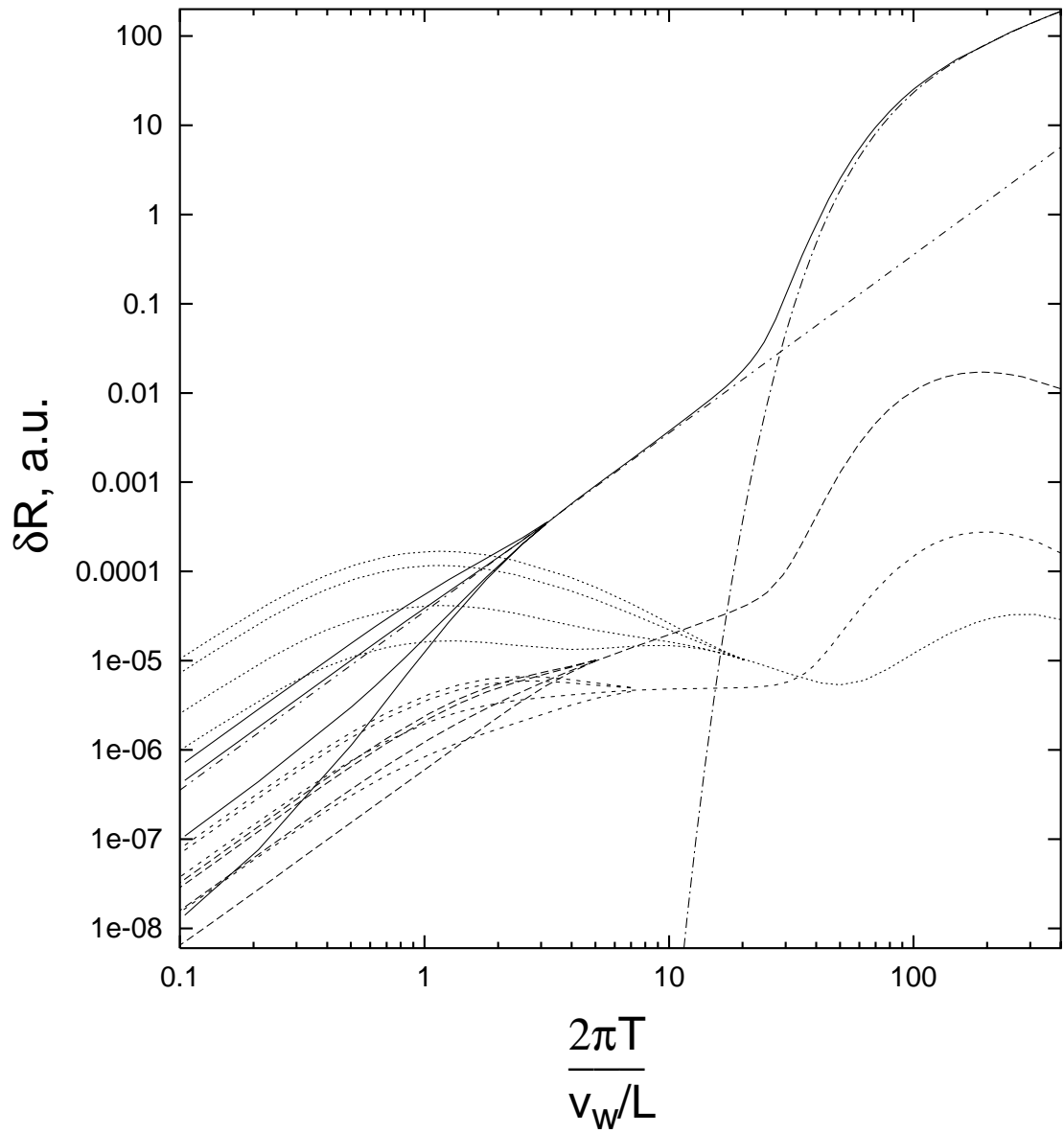
\end{document}